\newcommand{\gtappeq}{\raisebox{-0.6ex}{$\,\stackrel
{\raisebox{-.2ex}{$\textstyle >$}}{\sim}\,$}}

\documentclass[12pt,preprint]{aastex}
\usepackage{psfig}

\shorttitle{System parameters of DW~UMa}
\shortauthors{Araujo-Betancor et al. }

\begin{document}


\title{System Parameters of DW~Ursae Majoris\protect\footnote{Based on
observations with the NASA/ESA Hubble Space Telescope (HST), obtained
at the Space Telescope Science Institute (STScI), which is operated 
by the Association of Universities for Research in Astronomy, Inc., under NASA 
contract NAS5-26555, and with the Apache Point Observatory (APO) 3.5m
telescope, which is owned and operated by the Astrophysical Research
Consortium (ARC).}} 


\author{S. Araujo--Betancor, C. Knigge}
\affil{Department of Physics \& Astronomy. University of Southampton, 
Southampton~SO17~1BJ,~UK}

\author{K. S. Long}
\affil{Space Telescope Science Institute, Baltimore~MD 21218,~USA}

\author{D. W. Hoard}
\affil{SIRTF Science Center, IPAC, California Institute of Technology,
Pasadena~CA~91125, USA} 

\author{P. Szkody, B. Rodgers\protect\footnote{Present address: Gemini 
Observatory, AURA/Chile, P. O. Box 26732, Tucson~AZ~85726,~USA}~~\&
K. Krisciunas\protect\footnote{Present address: CTIO, 950 N. Cherry Ave, 
Tucson~AZ~85719,~USA}}
\affil{Department of Astronomy. University of Washington, Seattle~WA~98195,~USA}

\author{V. S. Dhillon} 
\affil{Department of Physics \& Astronomy. University of Sheffield,
Sheffield~S3~7RH,~UK}

\author{R. I. Hynes}
\affil{Department of Physics \& Astronomy. University of Southampton, 
Southampton~SO17~1BJ,~UK}

\author{J. Patterson \& J. Kemp\footnote{Also Joint Astronomy Centre,
University Park, 660 North A'ohoku Place, Hilo~HI 96720,~USA.}}
\affil{Department of Astronomy. Columbia University, New York~NY~10027,~USA}



\begin{abstract}
We present new constraints on the system parameters of the SW~Sextantis
star DW~Ursae~Majoris, based on ultraviolet ($UV$) eclipse
observations with the {\em Hubble Space Telescope}. Our data were 
obtained during a low state of the system, in which the $UV$ light 
was dominated by the hot white dwarf (WD) primary. The duration of the 
WD eclipse allows us to set a firm lower limit on the mass ratio, $q =
M_2/M_1 > 0.24$; if $q < 1.5$ (as expected on theoretical grounds) the
inclination must satisfy $i > 71^{\circ}$. We have also been able to
determine the duration of WD ingress and egress from our data. This
allows us to constrain the masses and radii of the system components
and the distance between them to be $0.67\leq M_1/M_{\odot} \leq
1.06$, $0.008\leq R_1/R_{\odot} \leq 0.014$, $M_2/M_{\odot} > 0.16$,
$R_2/R_{\odot} > 0.28$ and $a/R_{\odot} > 1.05$. If the secondary follows 
Smith \& Dhillon's mass-period relation for CV secondaries, 
our estimates for the system parameters become $M_1/M_{\odot} = 0.77
\pm 0.07$, $R_1/R_{\odot} = 0.012 \pm 0.001$, 
$M_2/M_{\odot} = 0.30 \pm 0.10$, $R_2/R_{\odot} = 0.34 \pm 0.04$, $q =
0.39 \pm 0.12$, $i = 82^{\circ} \pm 4^{\circ}$ and $a/R_{\odot} = 
1.14 \pm 0.06$. 

We have also obtained time-resolved $I$- and $K$-band photometry of DW~UMa
during the same low state. Using Bessell's spectral-type vs
$(I-K)$ color calibration, we estimate the spectral type of the donor
star to be $M3.5 \pm 1.0$. This latter result helps us to 
estimate the distance towards the system via Bailey's method as
$d = 930 \pm 160$~pc.

Finally, we have repeated Knigge et al.'s WD model atmosphere
fit to the low-state $UV$ spectrum of DW~UMa in order to account for the 
higher surface gravity indicated by our eclipse analysis. The best-fit
model with surface gravity fixed at $\log{g}=8$ has an effective
temperature of $T_{\rm eff} = 50,000 \pm 1000$~K. The normalization of the
fit also yields a second distance estimate, $d = 590 \pm 100$~pc. If
we adopt this distance and assume that the mid-eclipse $K$-band flux
is entirely due to the donor star, we obtain a second estimate for the
spectral type of the secondary in DW~UMa, $M7 \pm 2.0$. After discussing
potential sources of systematic errors in both methods, we conclude
that the true value for the distance and spectral type will probably 
be in between the values obtained by the two methods.   

\end{abstract}




\section{Introduction}
Nova-likes (NLs) are a subgroup of cataclysmic variables (CVs) in which
a late-type main sequence secondary loses mass onto a white
dwarf (WD) primary via Roche lobe overflow. If the WD does not have a strong
magnetic field, the transferred matter forms an accretion disk with a
bright spot where the stream of matter hits the disk. Unlike the more
commonly known dwarf nova type CVs, NLs do not undergo quasi-periodic
outbursts. They are instead characterized by a high, steady accretion
rate which prohibits the disk instability mechanism responsible for
the dwarf nova outbursts. However, some NLs with periods between 3-4~hrs also
exhibit low states during which mass transfer from the secondary and/or
accretion onto the WD decreases or shuts off completely. 
The reason for the low states is uncertain; one possibility is that they are 
related to magnetic activity of the secondary star (see Hessman,
G${\rm \ddot{a}}$nsicke \& Mattei 2000).
Supporting this theory is the fact that the orbital period of these
systems is very close to the upper edge of the CV period gap between 2-3~hrs. 
The absence of CVs in this region of the orbital period distribution,
is believed to occur as a consequence of a change in the internal
structure of the secondary. More specifically, it is thought that at
$P_{\rm orb} \approx 3$~hrs, the secondary becomes fully convective
and magnetic braking ceases to be the main mechanism by which angular
momentum is removed from the system. As a result, the secondary loses
contact with its Roche lobe and mass transfer ceases. Gravitational
radiation takes over as the dominant, but less efficient, angular
momentum loss mechanism. At $P_{\rm orb}\approx 2$~hrs 
the orbit has shrunk enough for the secondary to
re-establishes contact, and mass transfer resumes. For a thorough
review of CVs refer to Warner (1995). 

The subject of our paper, DW~Ursae~Majoris, is an eclipsing CV with a period
of $P_{\rm orb} =$ 3.28~hrs and belongs to a subclass of NLs called SW
Sextantis stars (Thorstensen et al.~1991). SW~Sex systems are grouped
together because they exhibit 
several unusual properties: (a) they are often 
eclipsing systems with orbital periods of 3-4~hrs; 
(b) their continuum eclipses are more V-shaped (as opposed to
U-shaped) than those of other NLs;
(c) their optical emission lines are single-peaked, instead 
of double-peaked (as expected for high-inclination, disk-formed lines); 
(d) their Balmer and HeI lines remain largely
unobscured during primary eclipse, but display absorption
events at the opposite orbital phase; (e) the radial velocity curves
derived from their optical emission lines lag substantially behind the
phase one expects from the WD on the basis of eclipses. Several models
have  been put forward to account for the SW~Sex stars,
and although each is capable of explaining a subset of the SW~Sex
behavior, none has so far been able to explain all the features listed
above (e.g. Knigge et al. 2000 and references therein).

One reason for our poor understanding of the SW~Sex phenomenon is the
scarcity of reliable system parameters for members of this class. This
scarcity is a direct consequence of the defining SW~Sex
characteristics. More specifically, radial velocity studies of SW~Sex
stars are of limited value, given the ubiquitous and significant phase
lags seen in the radial velocity curves of these systems. Eclipse studies
have been similarly unsuccessful, since WD contact points are not
evident in the V-shaped eclipse light curves of SW~Sex stars (perhaps
because the disks are self-occulting; Knigge et al. 2000). Also, the high
accretion rates exhibited by these systems during normal states 
cause the late-type main sequence secondary to be invisible against
the glare of the WD and/or accretion disk.

The goal of this paper is to improve on this situation by deriving a 
set of reliable system parameters for DW~UMa. This is possible because
a recent {\em Hubble Space Telescope} (HST) 
observation of the system found DW~UMa in a deep low state, during which the
WD dominated the 
ultraviolet ($UV$) light. This is the second, in a series of papers,
describing these observations. The first, Knigge et al. (2000),
dealt with the low-state $UV$ spectrum. Here we analyze
the time-resolved behavior around the eclipse and discuss $I$- and
$K$-band photometry obtained during the same low state. The remainder
of this article is organized as follows: In section 2, we describe the HST
and ground-based observations and their reduction. Next, in section 3,
we discuss our determination of the contact phases describing the
eclipse of the WD by the secondary. In section 4, we use the contact
phases an other information to determine the parameters of the binary
system and its constituents. Then, in section 5, we calculate
estimates for the distance to DW~UMa and for the spectral type of the 
donor star in two different ways: using our $I$- and $K$-band
photometry and re-fitting the Knigge et al.'s (2000) WD model
atmosphere to the low-state $UV$ spectrum of DW~UMa. Finally, 
in section 6, we discuss our results and compare the system parameters
we have derived to those previously reported for DW~UMa.

\section{Observations}
\label{obs}

The observations of DW~UMa with the HST {\em Space Telescope Imaging
Spectrograph} (STIS), took place on 1999 January 25~UT
and covered just over two complete cycles of DW~UMa's 3.28~hrs orbital
period. DW~UMa is in HST's continuous viewing zone (CVZ), so we
were able to observe the system almost continuously, with only 3 short
($\simeq 6$~minutes) interruptions between the four separate $UV$
exposures. The
instrumental set-up consisted of the $52\arcsec\times0.2\arcsec$ slit,
the FUV-MAMA detectors, and the G140L grating. This combination
provides a wavelength coverage of $1150-1720$~\AA~at a resolution of 
$\simeq1$~\AA~(FWHM). TIME-TAG mode was used throughout, so that  
individual photon arrival times were recorded at a sampling rate of
125 $\mu$s. 

Near-simultaneous optical photometry obtained from the
{\em MDM Observatory} on Kitt Peak, Arizona, puts DW~UMa 
at $V\simeq17.6$ around the time of the
HST observations. This indicates that the system was in a deep low
state at this time ($V \simeq 14.5$ in the high state). Given that the
accretion rate must be severely reduced during a low state, the hot WD
primary may be expected to dominate the $UV$ light in this
state. Indeed, Knigge et al. (2000) have already used the same data set
to show that the low-state, out-of-eclipse $UV$ spectrum of DW~UMa is
consistent with that of a hot WD. Our $UV$ eclipse observations
therefore provide us with an unusual opportunity to determine the WD
contact phases (and therefore system parameters) of this eclipsing 
SW~Sex star. 

For the purpose of the eclipse analysis, we constructed light
curves directly from the four TIMETAG files. These files contain a
list of the arrival times and detector positions of all recorded
photon events, and are therefore ideally suited for short-timescale
variability studies. Since we wanted to exclude geocoronal and other
strong lines from our light curves, we first determined an approximate
linear dispersion relation to relate pixel numbers in the dispersion
direction with physical wavelengths. Light curves were then
constructed at 1-s time resolution based on all counts recorded in
three continuum windows in the dispersion direction, {\em viz.} 1340 --
1380~\AA, 1410 -- 1520~\AA~and 1570 -- 1720~\AA~, and in a 0.36 arcseconds
window in the spatial direction. Since we are only 
interested in determining contact phases, the zero-level of the light
curves is irrelevant. We therefore did not carry out any background 
subtraction on our light curves. The background is, in any case, 
expected to be negligible: our continuum regions are well separated
from geocoronal lines, and dark current contributes only about 7 c/s 
across the entire FUV-MAMA detector, and therefore less than 0.1 c/s
in the small region used for extracting the count rate.

Figure~1 shows the resulting $UV$ light curve of DW~UMa. Several points
should be noted: (i) the count rate at mid-eclipse phases is close to
zero, confirming that there is no significant background contribution
in our light curves; (ii) there are sharp features in each of the
eclipses (especially eclipse 3), which are likely due to
ingress and egress of the WD; (iii) the light curves exhibit
substantial short-timescale, stochastic variability. 

In addition to the HST observations, we have obtained some 
(near-)infrared $I$- and $K$-band photometry of DW~UMa. The $I$-band
observations were taken using the MDM 1.3~m telescope on 1999 February
8 and 9 UT, roughly two weeks after the $UV$ HST observations. The infrared
$K$-band data were obtained on 1999 March 28 UT using the {\em Apache Point 
Observatory} (APO) 3.5~m telescope and the GRIMII
detector. Near-simultaneous $B$-band photometry (also obtained using
MDM) showed that DW~UMa was still in its low state during all of these
observing runs ($B>17$). All of the $I$- and $K$-band
observations were timed to coincide with eclipses of the system. The 
mid-eclipse magnitudes of DW~UMa in these filters were found to be
$I_{\rm kc} =$ 18.71 $\pm$ 0.14 (on the Kron-Cousins system) and
$K_{\rm ukirt} =$ 16.43 $\pm$ 0.14 (on the UKIRT system). These values
will be used in Section~5 to estimate the spectral type of the donor
star and to constrain the distance towards the system via Bailey's
(1981) method.

In order to convert the HJD times to orbital phases, we used the ephemeris 
\begin{equation}
T_{\rm mid-eclipse} = 2446229.00696(3) + 0.136606499(3) E,
\end{equation}
where the numbers in parentheses designate the errors on the last
digits. This is an updated version (Smith, Dhillon \& Marsh 2002) 
of the ephemeris presented by Dhillon, Jones \& Marsh
(1994). In practice, we still found an O-C shift of approximately
+0.0062 (73~s) with this new ephemeris. This is consistent with that
found by Smith et al. (2002) from a high state of DW~UMa observed at a
similar epoch to ours. Note that our O-C shift, which has been removed
in all of the phase-folded light curves shown below, could include a
contribution of a few seconds from the uncertainty in the STIS absolute timing (see Long 2000). 

\section{Method of Measuring the eclipse timings}

The phase intervals inside which the WD disappears (ingress) and
reappears (egress) from behind the occulting secondary are defined as
$\Delta\phi_{\rm wi}= \phi_{\rm w2}-\phi_{\rm w1}$ for ingress, and
$\Delta\phi_{\rm we}= \phi_{\rm w4}-\phi_{\rm w3}$ for egress, where
$\phi_{\rm w1}$,
$\phi_{\rm w2}$, $\phi_{\rm w3}$ and $\phi_{\rm w4}$ are the WD contact phases
(c.f. Figure~2). The half-flux
phases within each of these intervals, $\phi_{\rm wi}$ and $\phi_{\rm we}$
respectively, are the phases at which half of the light from the compact
object is eclipsed. The full-width at half out-of-eclipse intensity
is then defined as $\Delta\phi=\phi_{\rm we}-\phi_{\rm wi}$, or, in units of
time, $\Delta {\rm t} = P_{\rm orb} \Delta\phi$.

The WD contact phases were measured using the method described by Wood,
Irwin \& Pringle (1985). Figure 3 illustrates this procedure carried out
on eclipse 3. The light curve is smoothed with
a median filter of width 11~s. The amount of filtering chosen is a
compromise between the desire to reduce the noise (Poisson plus
intrinsic variability) and the need to retain the
most rapid changes in the eclipse. The smoothed light curve is
differentiated numerically to enhance regions of sharp brightness
variations. The derivative light curve is then smoothed
by applying a mean filter of width equal to the expected value for the
times of ingress and egress of the WD (43~s). The largest negative and
positive peaks after application of the mean filter indicate the
location of the mid-points of ingress and egress respectively. The
width at half-peak intensity of these features gives a rough estimate
of their duration. A spline function is fitted to the derivative
omitting those points inside the estimated ingress and egress
intervals. This last step is carried out to account for the
contribution (if any) of the extended and slowly varying eclipse of
the disk. The fitted function is then subtracted from the derivative, 
and the result is smoothed with a median filter of width equal to the
median filter used in the first part of the reduction. The
contact phases are located by identifying the points where the
resultant derivative departs from zero.

The main difficulty in identifying the WD contact phases is the
presence of stochastic variability on timescales that are only
slightly longer than the duration of WD ingress/egress. Figures~1 and
2 show that this type of variability severely limits our ability to
determine WD contact phases for eclipses 1 and 2, in particular. We
therefore did not use these two eclipses for measuring the contact
points, but only to provide qualitative checks on the contact phases
determined from eclipse 3. Figure~2 confirms that the contact phases
derived from eclipse 3 are indeed consistent with eclipses 1 and 2. 
The final value for the duration of the ingress/egress of the WD is
$\Delta\phi_{\rm wd}=(\Delta\phi_{\rm wi}+\Delta\phi_{\rm we})/2$. In units
of time, this is $\Delta {\rm t}_{\rm wd} = P_{\rm orb} \Delta\phi_{\rm wd}$. Our final estimates
of $\Delta {\rm t}$ and $\Delta {\rm t}_{\rm wd}$ are listed in the first section of
Table~1.

\section{Binary parameters}

The mass ratio, $q=M_2/M_1$ ($M_1$ and $M_2$ are the masses of the WD
and secondary, respectively), and inclination, $i$, of the system can
be constrained by considering the eclipse of a point source 
by a Roche-lobe-filling secondary. More specifically, the observed
eclipse duration ($\Delta \phi$) defines a unique relation between $q$
and $i$. We calculated this relation by using a method similar to Chanan,
Middleditch \& Nelson (1976). Following their Figure 1, we define a spherical
coordinate system centered on the WD, in which the direction vector
($\theta$,$\phi$)~=~(90$^{\circ}$,0$^{\circ}$) points towards the 
centre of the secondary. The main idea is to consider the intersection
of the secondary's Roche lobe and the plane $\phi=\Delta \phi /
2$. This reduces the problem to finding the minimum value of $\theta$
along the curve defined by this intersection. For each $q$, this
minimum value is then just the inclination $i$ that produces an
eclipse of width $\Delta \phi$.  The method requires the critical
value of the Roche potential. This is calculated by using the fact
that the net effective force vanishes at the inner Lagrangian point;
 $\partial\psi/\partial r=0$ at $L_1$, where $\psi$~is the
effective potential. The $q$ vs $i$ relation for our measured $\Delta \phi$
is shown in Figure~4.

The mean duration of the ingress and egress features of the eclipse
($\Delta\phi_{\rm wd}$), together with the calculated numerical relation
$i=f(q,\Delta\phi)$, can be used to constrain the remaining
binary parameters. In practice, we use the full Roche lobe geometry of
the secondary to obtain the radius of the spherical compact primary. For
each (i, q) pair, our algorithm computes the radius of the WD scaled
to the binary separation, $R_1/a$, that successfully reproduces the observed value for 
$\Delta\phi_{\rm wd}$. The WD is assumed to be fully
visible; this is appropriate here, since there is no evidence for an optically
thick disk in the low state.
 
Kepler's third law, which defines the binary separation, $a$,
\begin{equation}
a = \left(\frac{G}{4 \pi^2}\right)^{1/3} M_1^{1/3} \left(1 +
q\right)^{1/3} P_{\rm orb}^{2/3},
\end{equation} 
is then combined with a theoretical WD mass-radius relation to yield a
definite WD mass, $M_1$, for each $q$. The inset in Figure~5 shows the
resulting constraints on the component masses for a wide range of mass ratios.

Finally, we use Eggleton's (1983) approximation
\begin{equation}
\frac{R_2}{a} = \frac{0.49 q^{2/3}}{0.6 q^{2/3} + \ln \left(1 +
q^{1/3}\right)}. 
\end{equation}
to estimate the volume-averaged radius of the secondary, $R_2$.

The theoretical WD mass-radius relation that we use was taken from 
Panei, Althaus \& Benvenuto (2000) and describes hot ($T_{\rm eff} =
50,000$~K) CO-core WDs with a relatively massive, non-degenerate
hydrogen envelope ($M_{\rm env}/M_{1} = 10^{-5}$). The effective
temperature adopted here is close to the value (46,000~K)
inferred by Knigge et al. (2000) from a WD model atmosphere fit.
For reference, the spectrum analyzed by Knigge et al. (2000) was
constructed from the time interval between eclipses 2 and 3, when the
count rate was high and relatively stable. We used a slightly higher
temperature here, because our analysis indicates a higher surface
gravity ($\log g$). More specifically, preliminary parameters inferred
with a $T_{\rm eff}=46,000$~K mass-radius relation yielded WD
parameters corresponding to $\log g \simeq 8$. As discusses in more
detail in Section~5, we then refit the low-state UV spectrum with this
new gravity estimate and obtained a new temperature estimate of
$T_{\rm eff} = 50,000$~K. We note that the systematic errors
associated with the presence or absence of a hydrogen envelope and
with errors in temperature up to at least 10,000~K are smaller than 
the random errors resulting from observational uncertainties. Hence,
the iteration process does not change the parameter values
substantially, but merely improves their internal consistency. 

At this stage, we can already place useful constraints on most of the
system parameters, which are listed in the second section of
Table~1. These constraints are only based on the eclipse geometry and
the theoretical WD mass-radius relationship and are therefore
particularly robust. If the mass ratio is made to satisfy $q < 1.5$
(see Figure~5), as
required for stable mass-transfer from any main-sequence secondary
(Politano 1996; de Kool 1992), the constraints are tightened, as shown
in the third section of Table~1.

If we are willing to assume that the secondary star is ``typical'' for
a CV with DW~UMa's orbital period, we can tighten our constraints even
further by using Smith \& Dhillon's (1998) mass-period relationship
for CV secondaries. The resulting estimate for the mass of DW~UMa's
secondary is $M_2/M_{\odot} = 0.3 \pm 0.1$, where the error is 
based on the rms scatter quoted by Smith \& Dhillon (1998) for this
relationship. The effect of this $M_2$ constraint on $M_1$ is shown in 
Figure~5. The corresponding estimates for all system
parameters are listed in the fourth section of Table~1.  

\section{Spectral Type and Distance}
\label{SpT-Dist}

Distances for CVs can be estimated via Bailey's (1981) method. This is
based on the fact that the $K$-band surface brightness, $S_{\rm K}$, is only
a weak function of spectral type among isolated, late-type,
main-sequence stars. The definition of $S_{\rm K}$ is 
\begin{equation}
S_{\rm K}= K+5-5\log{d} + 5\log{R_2}
\label{sk1}
\end{equation}
where $K$ is the star's $K$-band magnitude, $R_2$ is its radius (in
solar units) and $d$ is its distance (in parsecs). The most recent
calibration of $S_{\rm K}$ against $(V-K)$ color for late-type stars is due to
Beuermann (2000) 
\begin{equation}
S_{\rm K} = 2.98 + 0.264 (V - K_{\rm cit}).
\label{sk2}
\end{equation} 
In this equation, V is the usual Johnson waveband, and $K$ is on the
CIT system. Our own $K$-band observations are on the UKIRT system, but 
according to Leggett (1992), $K_{\rm cit} = K_{\rm ukirt}$. By
combining Equations~4~and~5, the distance towards a
CV can be determined.

In the case of DW~UMa, we have an estimate of the mid-eclipse $K$-band 
magnitude obtained during a low-state, i.e., at a
time when any WD and disk contribution should be relatively
small. This estimate -- $K_{\rm ukirt} = 16.43 \pm 0.14$
(Section~2) -- should therefore be quite close to the true $K$-band 
magnitude of the secondary.

In order to determine, $S_{\rm K}$, one needs the $(V-K)$ color of the
secondary. In practice, this is difficult to obtain in most CVs,
including DW UMa, because the $V$-band is usually dominated by the WD
and/or the accretion disk of the system, and hence the $(V-K)$ color
must be estimated from its (known/assumed) spectral type. In our case,
although we do not have a $V$ magnitude for the secondary, we do have
an estimate -- $I_{\rm kc} = 18.71 \pm 0.14$ -- of
the mid-eclipse $I$-band magnitude in the low state. We can therefore 
use the observed $(I-K)$ color to estimate the spectral type of the
secondary and use this to predict its $(V-K)$ color. 

More specifically, we use the spectral-type vs color calibration
presented by Bessell (1991). This calibration uses optical colors 
(BVRI) on the Kron-Cousins system, but infrared colors (JHKL) on
Bessell \& Brett's (1988) homogenized system. We therefore used the 
transformations provided by Bessell \& Brett (1988) to transform our
$K_{\rm ukirt} = K_{\rm cit}$ to $K_{\rm BB}$ (on the Bessell 
\& Brett system), although the resulting correction turned out to be
insignificant (+0.02). Armed with the $(I_{\rm kc}-K_{\rm BB})$ color of the secondary 
(Table~2), we then interpolated on Table~2 of Bessell (1991) to estimate the spectral
type of the secondary in DW~UMa, $M3.5 \pm 1.0$ (to the nearest
half-subtype). This may be compared with the
spectral type of $M4.2 \pm 0.8$ predicted by Smith \&
Dhillon's (1998) orbital period-spectral type relation for CV
secondaries (the error here corresponds to the rms scatter of the data
about their relation). For this spectral type, Bessell (1991) predicts 
$(V-K) = 5.0 \pm 0.7$, which in turn gives $S_{\rm K} = 4.3 \pm
0.2$. Here, we have dropped the subscript on $K$, since the differences
between the various systems are insignificant compared to the
uncertainty on the color. 

Having determined both $K$ and $S_{\rm K}$, we can now estimate the
distance to DW~UMa. We use the secondary's radius derived 
from the Smith \& Dhillon (1998) mass-period relation for CV
secondaries and Eggleton's (1983) approximation for the radius,
$R_2/R_{\odot} = 0.34 \pm 0.04$ (Section~4). Solving Equation~4 
for $d$ finally yields an estimate $d = 930 \pm 160$~pc (Table~2).   

By comparison, Knigge et al. (2000) estimated $d = 830 \pm 150$~pc
from the normalization of their fit to the low state $UV$ spectrum and
the radius of the WD inferred from their $\log{g}$. This agrees very
well with our own estimate, but this is partly fortuitous since their fit
yielded $\log g = 7.60$, whereas our new parameters suggest $\log
g\simeq 8$. We have therefore refit the low state $UV$ spectrum with
surface gravity fixed at  $\log g = 8$. Visually, this new fit is
almost indistinguishable from the fit shown in Knigge et al. (2000),
with a reduced Chi-squared of $\chi_{\nu}^2 = 1.55$ (up from
$\chi_{\nu}^2 = 1.50$). The parameters describing this new fit are
listed in Table~3. Combining this with our new estimate of $R_{1}/R_{\odot} =
0.012 \pm 0.001$, we obtain a new value for the distance of
$d = 590 \pm 100$~pc. If we adopt this distance estimate 
and assume that the mid-eclipse $K$-band flux is entirely due to the
secondary, we can also obtain a new estimate for the spectral type of
the donor star. With $d = 590 \pm 100$~pc, $K_{\rm ukirt} = K_{\rm cit} =
16.43 \pm 0.14$ and $R_2/R_{\odot} = 0.34 \pm 0.04$, Equations~4 and 5
predict $(V-K) = 8.5 \pm 1.9$. For this color, Bessell (1991) predicts
$SpT_2=M7 \pm 2.0$ (to the nearest half-subtype). If not all of the
$K$-band light is due to the secondary, the spectral type must be even
later. 

Our two sets of distance and spectral type estimates are only
marginally consistent with each other. We therefore briefly consider 
potential sources of systematic errors in both of the methods we 
used to obtain them. Most distance estimates based on Bailey's method
are, strictly  
speaking, lower limits, since the observed $K$-band magnitude may  
contain residual disk and/or WD contributions (Hoard et al. 2002). 
In our case, the situation is less clear. This is because our value
for $d$ depends on our spectral type estimate, which in turn is 
based on the observed $(I-K)$ color. But contamination by sources
other than the secondary will affect the measured $I$- and $K$-band 
magnitudes differently. Indeed, any reasonable contaminating spectrum
will be bluer than that of the secondary, in which case the I-band
would be more contaminated than the K-band. If so, then our measured
$(I-K)$ 
color would be too blue, our spectral type too early and our predicted
$(V-K)$ color also too blue. Thus this effect may cause the distance to
be {\em over-}estimated. It is not clear if our estimate for $d$ is
affected by these biases, and, if so, which of them dominates.

By contrast, the key assumption in the WD fit is that the observed
spectrum is due entirely to a fully visible WD. We first consider the
possibility that the observed spectrum actually includes a 
contribution from another source. If the spectrum of this source is 
redder than that of the WD, then the true WD spectrum must be (a) bluer
and (b) fainter than the observed spectrum. Effect (a) would tend to
make our distance estimate too low, because a hotter WD is intrinsically
brighter and, for given radius and observed flux, must therefore be
further away. Effect (b) would also lead us to underestimate the
distance, since the latter scales with observed flux as
$F_{\rm obs}^{-1/2}$. Similarly, if the contaminating spectrum is bluer
than the WD spectrum (which seems unlikely), then effect (a) above is
reversed and would compete with effect (b). Finally, if a significant
portion of the WD is obscured, the observed spectrum arises from a
projected area that is smaller than $\pi R_{1}^2$. For a given
observed flux (and hence a fixed fit normalization $N = 4 \pi
R_1^2/d^2$), this would imply that our distance estimate is too large. 

In principle, we consider the distance estimate based on the WD model
fit more direct, and therefore more reliable, than the estimate
derived from Bailey's method. However, the spectral type of the
secondary we infer from the former estimate seems suspiciously too
late by comparison to other CVs with similar orbital periods (Beuermann et
al. 1998). On balance, we therefore expect that the true distance and
spectral type will turn out to lie between the values suggested by the 
two methods. 

\section{Discussion \& Conclusions}

SW~Sex stars display a range of peculiarities that do not seem to fit
the standard steady accretion disk model for NLs. There is currently
no single agreed-upon explanation for the SW~Sex phenomenon. Indeed,
no real consensus has been reached about whether SW~Sex stars deserve
a NL-sub-class label in the first place. Rather, the SW~Sex {\em
syndrome} is so widespread (also seen in X-ray binaries; Hynes et
al. 2001) that we must consider the possibility that
some important element is missing in our standard picture of the
accretion flows. Whatever the nature of the SW~Sex stars, they may 
play an important role in CV evolution: with very few exceptions
(e.g., BT Mon with $P_{\rm orb}=8$~hrs; Smith, Dhillon \& Marsh 
1998), SW~Sex stars have orbital periods falling in the 3 -- 4 hrs
range, just above the period gap.

Reliable system parameters are desperately needed in order to
understand the origin of the SW~Sex phenomenon. Unfortunately, the 
very characteristics that distinguish SW~Sex stars from
other NLs have also prevented us from accurately determining their
physical and geometrical parameters. In particular, the
phase-shifted radial velocity curves seen in these systems make it
very difficult to accurately determine the $K$-velocities of the
component stars (from which other system parameters could then
follow). In addition, the high accretion rates exhibited by the SW~Sex
stars hamper the detection of the individual components of the system 
since the accretion disk dominates the emission and may even be
self-occulting (Knigge et al. 2000). These latter effects are greatly
reduced during the sporadic low states displayed by at least some of
the SW~Sex stars. 

In the case of DW~UMa, radial velocity studies have been carried out
based on both high-state (Shafter, Hessman \& Zhang 1988) and
low-state (Dhillon at al. 1994; also see Rutten \& Dhillon 1994) 
observations. However, as emphasized by both sets of authors,
neither study has provided reliable results. In the former study, 
significant ($55^{\circ}-75^{\circ}$) phase lags were seen in all the radial
velocity curves from which $K_1$ was estimated; in the latter study,
the radial velocity curve was based on emission lines that clearly
arise on the irradiated front face of the secondary (Rutten \& Dhillon
1994), and may therefore provide only a lower limit for $K_2$.

Our HST low state $UV$ observations of DW~UMa have provided a
rare opportunity to accurately determine the system parameters of an
SW~Sex star from the eclipses of its WD. This has been possible because
the accretion disk contribution is dramatically reduced, revealing the 
sharp ingress and egress features that mark the WD eclipse. Therefore, we
have been able to avoid many of the difficulties associated with
radial velocity studies. However, potential sources of systematic
errors remain and include (1) our assumption that the WD is entirely
unobscured, (2) the application of the mass-period relation for CV
secondaries, and (3) the application of the theoretical mass-radius
relation for isolated WDs.  

As it turns out, the system parameters listed in Table~1 agree
reasonably well with the constraints inferred by Shafter et al. (1988) 
and Dhillon et al. (1994) from radial velocity
analyses. However, as argued above, we consider our new estimates to be 
considerably more reliable, particularly the ones labelled
``fundamental'' in Table~1. Our estimate
of the WD mass ($M_1/M_{\odot} \simeq 0.77$) is considerably higher
than that obtained by Knigge et al. (2000) from their model atmosphere
fit to the low state HST $UV$ spectrum ($M_1/M_{\odot} \simeq 0.5$). Their estimate
was based on the surface gravity inferred from the spectral fit and
essentially the same mass-radius relation used here but for a WD with
$T_{\rm eff}= 46,000$~K (Panei et al. 2000). 
Given the systematic uncertainties inherent in spectroscopic 
$\log{g}$ estimates, we consider our new set of constraints on the 
WD parameters more reliable. 

Both of our distance estimates, $d = 930 \pm 160$~pc (from Bailey's
method) and $d = 590 \pm 100$~pc (from the WD model fit), indicate
that DW~UMa is quite far away. For comparison, Marsh \& Dhillon
(1997) estimated $d \gtappeq 850$~[$450$]~pc if the secondary is an
$M4$~$[M5]$ star, based on the absence of clear secondary signatures
in their low-state $I$-band spectroscopy. We note, however, that even
though our estimate of $I \simeq 18.7$ for the secondary star is
fainter than the (out-of-eclipse) $I$-band magnitude they actually
observed ($I \simeq 18$), it is brighter than the limit of $I > 19.5$
($<25\%$) they suggest for the secondary's contribution to their
spectrum. If the mid-eclipse $I$-band flux in our date is due to the
secondary, the latter should have contributed roughly half of the
$I$-band flux in their spectrum. 

We conclude this paper by reiterating the need for accurate system
parameters for the SW~Sex stars. Given that most SW~Sex stars are
high-inclination systems, it is natural to try and exploit this in the
way we have done here. However, eclipse analyses based on WD ingress
and egress features are usually impossible, since the light from these
systems is usually dominated by their (possibly self-occulting) accretion
disks. In the case of DW~UMa, we have succeeded only because the
system was caught in a deep low state. This raises the obvious
question whether other SW~Sex systems may also exhibit such low
states. This would obviously be interesting for its own sake, but
would also open a new avenue of attack for determining their system
parameters.  

A partial answer to this question is already available: of the 14 
objects listed as SW~Sex stars in the G\"{o}ttingen Online CV
Catalog\footnote{http://www.cvcat.org}, 5 (including DW~UMa) are
already classified as VY~Scl stars (i.e., nova-like variables that
exhibit occasional low states). We therefore advocate a long-term
photometric monitoring program of SW~Sex stars. This would tell us 
whether all SW~Sex stars are also VY~Scl stars and permit us to 
exploit the low states when they occur.

\acknowledgments
We are grateful to Leandro Althaus for providing his WD mass-radius 
relations in electronic form. Support for this work was provided by NASA
through grant GO-7362 from the Space Telescope Science Institute
(STScI), which is operated by AURA, Inc., under NASA contract
NAS5-26555. R.I. Hynes acknowledges support from grant F/00-180/A from 
the Leverhulme Trust. We would like to thank T. R. Marsh for providing
a code for the full Roche lobe geometry analysis. We also wish to
thank the referee, J.Thorstensen, for his insightful comments and suggestions.

\newpage




\newpage

\footnotesize
\protect\begin{deluxetable}{lc}
\tablewidth{200pt}
\tablecaption{Parameters inferred from the eclipse analysis of DW~UMa.\label{tbl-wd1}}
\tablehead{
\colhead{Parameter}& 
\colhead{Value}}
\startdata
\multicolumn{2}{c}{Eclipse measurements} \nl \hline
$\Delta {\rm t}$~($s$) & 969 $\pm$ 4~ \nl
$\Delta {\rm t}_{\rm wd}$~($s$) & 48 $\pm$ 3 \nl \hline \nl
\multicolumn{2}{c}{Fundamental constraints} \nl \hline
$q$~($M_2/M_1$) & $>$ 0.24  \nl
$M_1$~($M_{\odot}$) & 0.67 $\leq M_1 \leq$ 1.06 \nl
$R_1$~($R_{\odot}$) & 0.008 $\leq R_1 \leq$ 0.014 \nl
$M_2$~($M_{\odot}$) & $>$ 0.16  \nl
$R_2$~($R_{\odot}$) & $>$ 0.28  \nl
$a$~($R_{\odot}$)   & $>$ 1.05  \nl \hline \nl
\multicolumn{2}{c}{With $q<1.5$} \nl \hline
$q$~($M_2/M_1$)     & 0.24  $\leq q \leq$ 1.5  \nl
$M_1$~($M_{\odot}$) & 0.67  $\leq M_1 \leq$ 0.96 \nl
$R_1$~($R_{\odot}$) & 0.009 $\leq R_1 \leq$ 0.014 \nl
$M_2$~($M_{\odot}$) & 0.16  $\leq M_2 \leq$ 1.44 \nl
$R_2$~($R_{\odot}$) & 0.28  $\leq R_2 \leq$ 0.62 \nl
$a$~($R_{\odot}$)   & 1.05  $\leq a   \leq$ 1.50 \nl 
$i$~(degrees)       & $>$ 71 \nl \hline \nl
\multicolumn{2}{c}{With $M_2$-$P_{\rm orb}$ relation} \nl\hline  
$q$~($M_2/M_1$)     & 0.39  $\pm$ 0.12  \nl
$M_1$~($M_{\odot}$) & 0.77  $\pm$ 0.07  \nl
$R_1$~($R_{\odot}$) & 0.012 $\pm$ 0.001 \nl
$M_2$~($M_{\odot}$) & 0.30  $\pm$ 0.10  \nl
$R_2$~($R_{\odot}$) & 0.34  $\pm$ 0.04 \nl 
$a$~($R_{\odot}$)   & 1.14  $\pm$ 0.06 \nl 
$i$~(degrees)       & 82    $\pm$ 4    \nl 
\enddata
\protect\end{deluxetable}
\normalsize

\newpage
\pagebreak

\footnotesize
\protect\begin{deluxetable}{lc}
\tablewidth{150pt}
\tablecaption{Parameters inferred from $I$- and $K$-band photometry.\label{tbl-ik}}
\tablehead{
\colhead{Parameter}& 
\colhead{Value}}
\startdata
$(I_{\rm kc}-K_{\rm BB})$	& 2.3 $\pm$ 0.2   \nl
$Spectral~Type$		& $M3.5 \pm 1.0$ \nl 
$d$~(pc) 		& 930 $\pm$ 160 \nl 
\enddata
\protect\end{deluxetable}
\normalsize

\newpage

\footnotesize
\protect\begin{deluxetable}{lc}
\tablewidth{300pt}
\tablecaption {Parameters inferred from the WD model atmosphere fit.\label{tbl-wd2}}
\tablehead{
\colhead{Parameter}& 
\colhead{Value\tablenotemark{a}}}
\startdata
$T_{\rm eff}$~(K) & 50,000 $\pm$ 1,000~ \nl
$\log{g}$~($\log{}$cm~s$^{-2}$) & 8\tablenotemark{b} \nl
$v\sin{i}$~(km~s$^{-1}$) & 345 $\pm$ 86 \nl
$Z$~($Z_{\odot}$) & 0.71 $\pm$ 0.14 \nl
$N=4\pi (R_1/d)^2$~(10$^{-24}$) & 2.68 $\pm$ 0.10 (stat) $\pm$ 0.67
(sys)\tablenotemark{c} \nl
$E_{\rm B-V}$~(mag) & 0.004 $\pm$ 0.004 \nl
$\log{N_{\rm H}}$~($cm^{-2}$) & 19.4 $\pm$ 0.06~ \nl
$d$~(pc) & 590 $\pm$ 100 \nl
$Spectral~Type$ & $M7 \pm 2.0$\tablenotemark{d} \nl
\enddata
\tablenotetext{a}{Uncertainties on the fit parameters correspond to $2
\sigma$ confidence level for one interesting parameter}
\tablenotetext{b}{Fixed value}
\tablenotetext{c}{The systematic error accounts for the uncertainty in
the absolute flux calibration of the spectrum and and in the
normalization of the models (Brown, Ferguson \& Davidsen 1996; Knigge
et al. 1997)} 
\tablenotetext{d}{This value was calculated assuming the mid-eclipse
K-band flux is entirely due to the secondary and adopting 
$d= 590 \pm 100$ and $R_2/R_{\odot}=0.34 \pm 0.04$}
\protect\end{deluxetable}
\normalsize

\clearpage

\protect\begin{figure*}[t]
\label{lc}
\vspace{-2cm}
\plotone{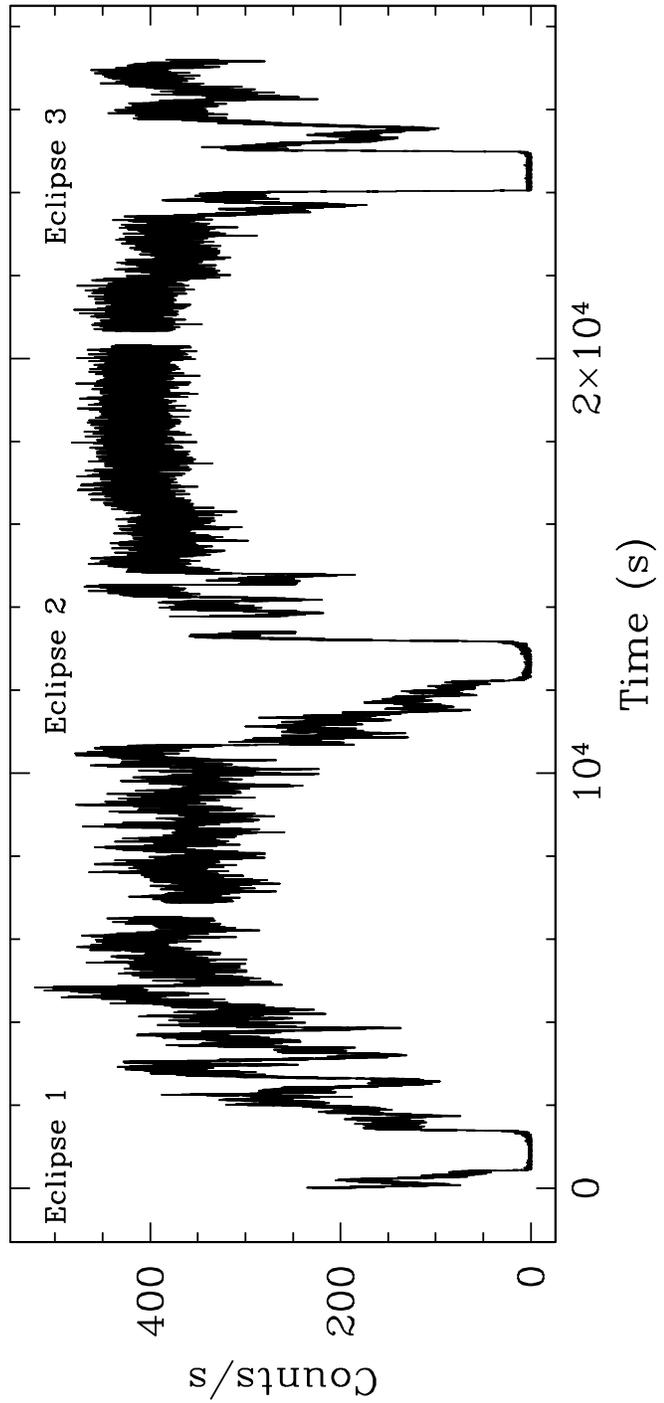}
\figcaption{DW~UMa light curve, with 1~s time resolution, constructed
from 3 continuum windows (1340 -- 1380~\AA, 1410 -- 1520~\AA~and~1570
-- 1720~\AA) of the HST $UV$ spectra. The start time corresponds to 
2451203.933 HJD and the total duration of the light curve is 7.55~hrs.}
\protect\end{figure*}

\clearpage

\protect\begin{figure*}[t]
\label{eclipses}
\plotone{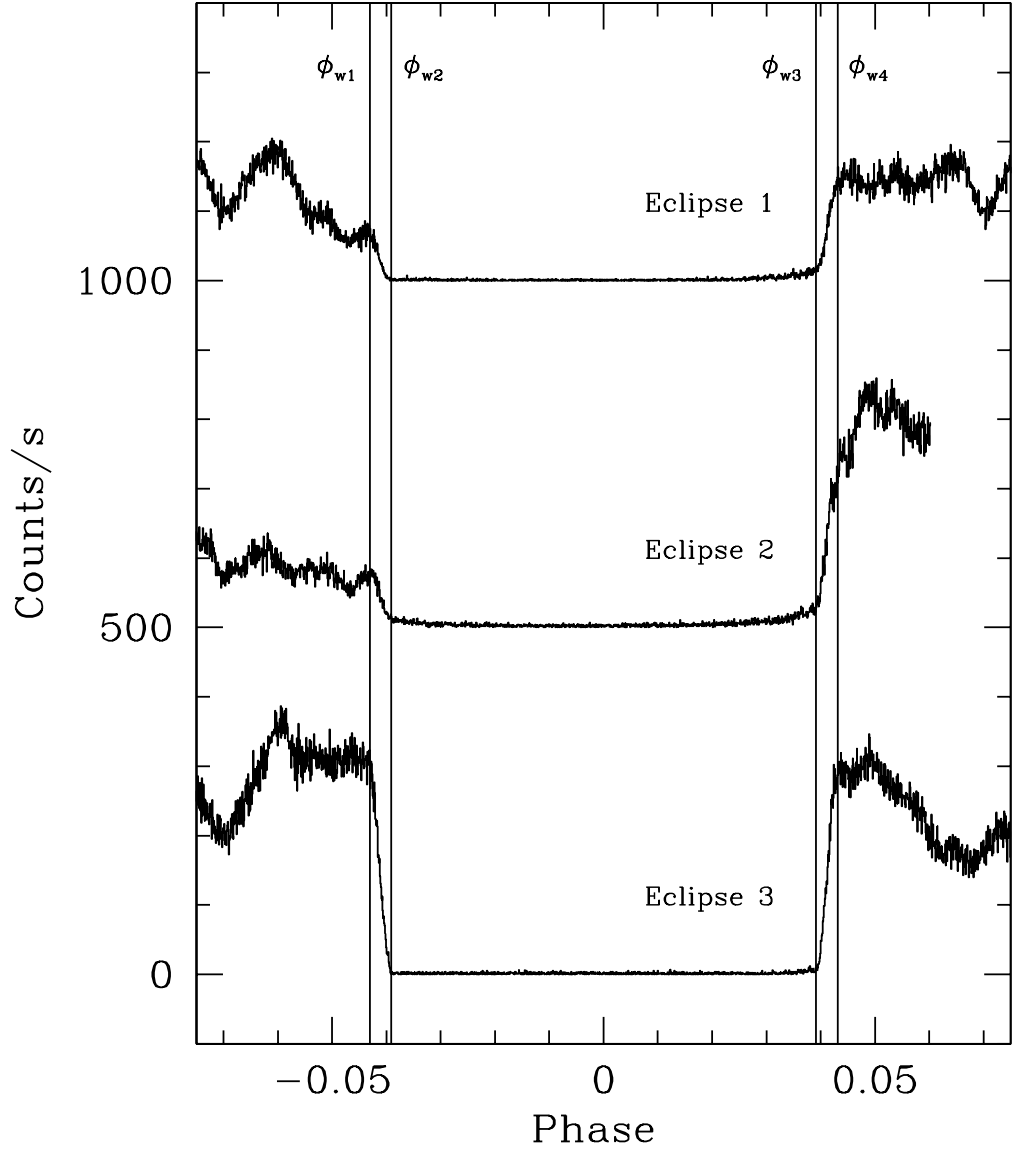}
\figcaption{The three individual eclipses of Figure 1 plotted against
orbital phase. Eclipse 2 and 1 are offset upward by 500 and 1000
counts/s respectively. The vertical lines mark the locations of the
contact phases ($\phi_{\rm w1}$, $\phi_{\rm w2}$, $\phi_{\rm w3}$ and
$\phi_{\rm w4}$) inferred
from eclipse 3 (see Figure~3).}
\protect\end{figure*}

\clearpage

\protect\begin{figure*}[t]
\label{phases}
\plotone{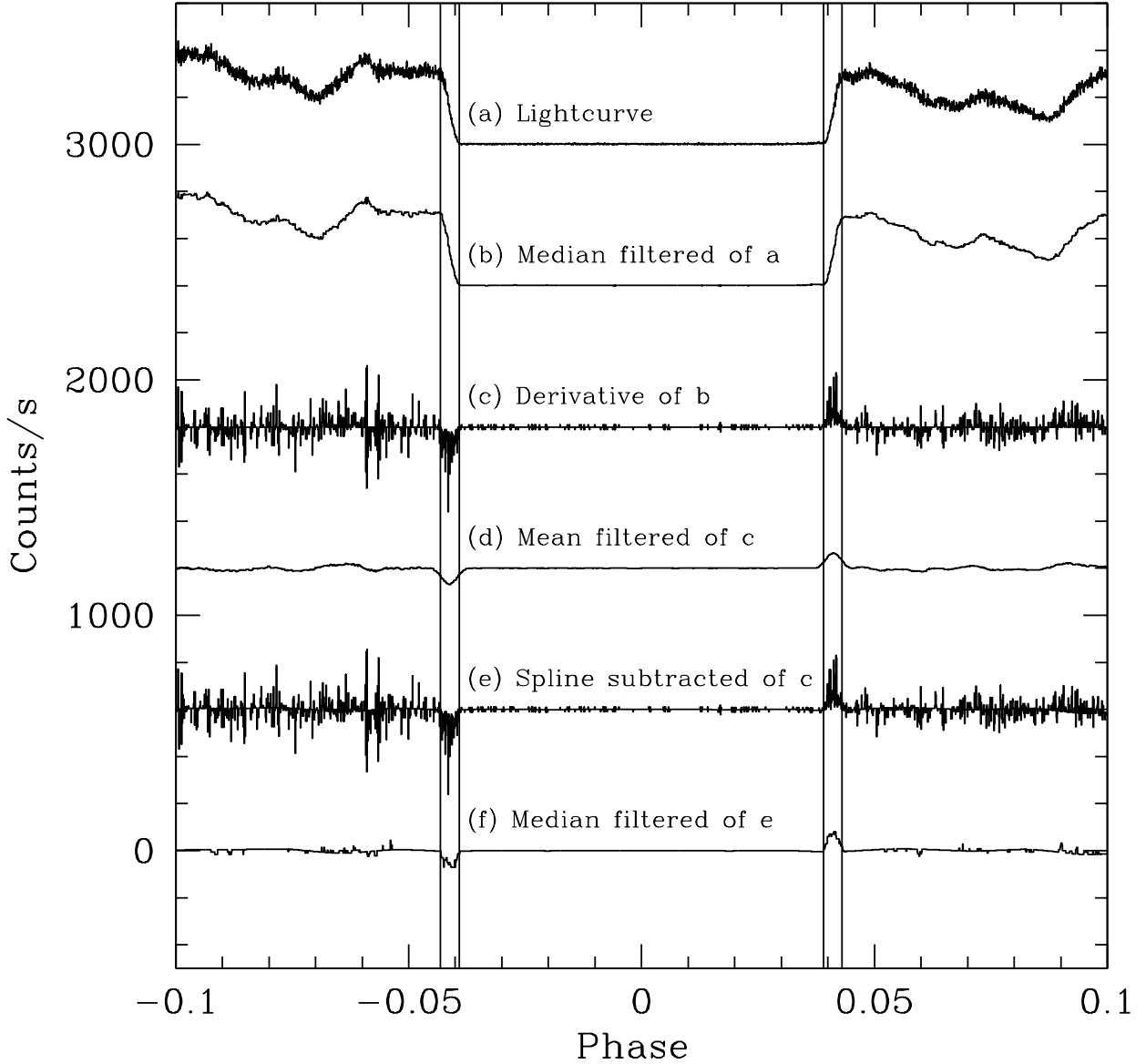}
\figcaption{Illustration of the method used to obtain the contact
phases from eclipse 3 of the light curve of DW~UMa. Curves (a) - (e)
were successively displaced upward by 600 counts/s each from curve
(f). Curves (c), (d), (e) and (f) have been multiplied by a
factor of 10~s for clarity. (a) Original light curve with 1~s time
resolution. (b) Median-filtered light curve (filter width 11~s). (c)
Derivative of the median-filtered light curve. (d) Mean-filtered
derivative (filter width 43~s). The highest negative and positive points in
this curve indicate the location of the mid-points of ingress and egress
respectively, (i.e., $\phi_{\rm wi}$ and $\phi_{\rm we}$). (e) Result of the
spline fit and subtraction from the derivative light curve. (f) Median
filtered version of (e) (filter width 11~s). The vertical lines
represent the measured positions of the contact points $\phi_{\rm
w1}$, $\phi_{\rm w2}$, $\phi_{\rm w3}$ and $\phi_{\rm w4}$.}
\protect\end{figure*}

\clearpage

\protect\begin{figure*}[t]
\label{qandi}
\plotone{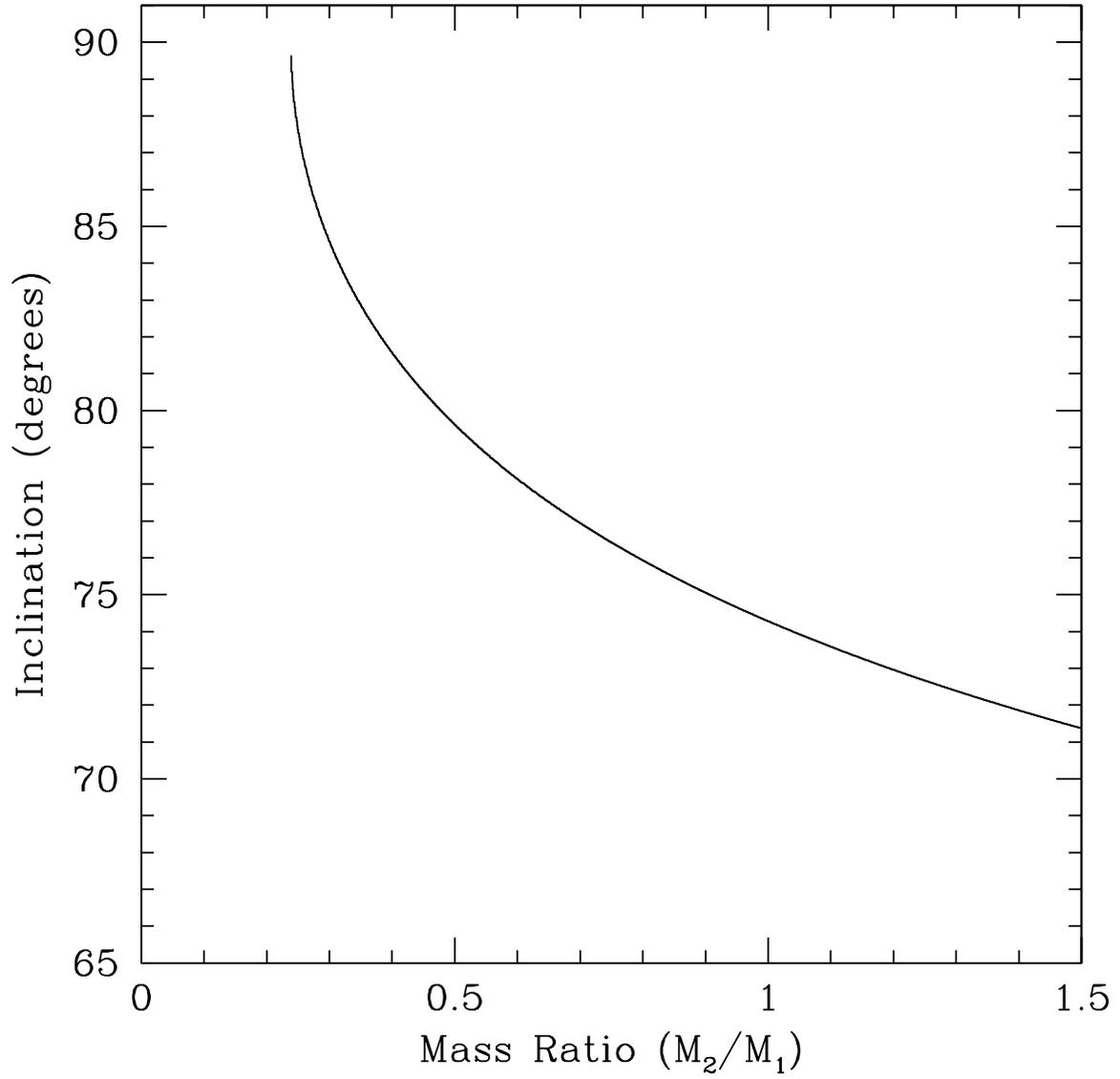}
\figcaption{Orbital inclination, $i$, as a function of mass ratio,
$q$. Note that the uncertainty limits on the $q$ vs $i$ relation were 
not plotted as they were negligible compared to the scale of figure.}
\protect\end{figure*}

\clearpage

\protect\begin{figure*}[t]
\label{massradius}
\plotone{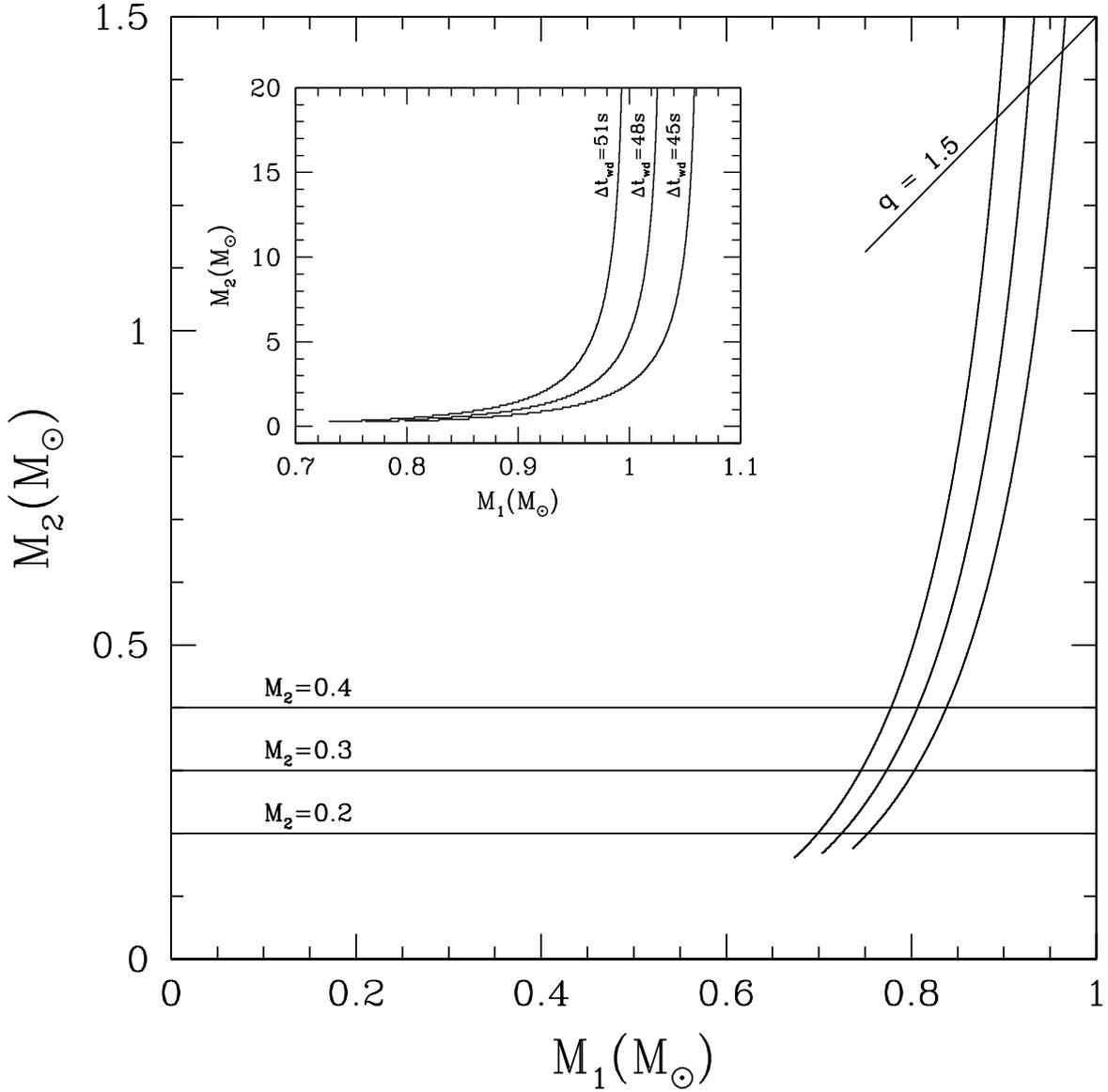}
\figcaption{Constraints on the component masses of DW~UMa. The three
curves in both plots correspond to $\Delta {\rm t}_{\rm wd}=$ 45~s, 48~s and
51~s. The inset
shows a larger range of values for $M_2$ so that the fundamental upper limits on $M_1$
can be clearly seen. The $q=1.5$ line on the main plot shows the upper
limits on the component masses based on theoretical grounds (see
text). The horizontal lines represent 
the value of $M_2/M_{\odot} = 0.3 \pm 0.1$ obtained from the $M_2-P_{\rm orb}$ 
relation of Smith \& Dhillon (1998). The corresponding value for the
mass of the WD is $M_1/M_{\odot}= 0.77 \pm 0.07$.}
\protect\end{figure*}


\end{document}